\documentstyle[prb,twocolumn,aps]{revtex}
\begin{document}
\draft
\title{Schwoebel barriers on Si(111) steps and kinks}
\author{S.\ Kodiyalam, K.E.\ Khor and S.\ Das Sarma}
\address{Department of Physics, University of Maryland, College Park, Maryland 
20742-4111}
\address{\mbox{ }}
\address{\parbox{16cm}{\rm \mbox{ }\mbox{ }\mbox{ }
Motivated by our previous work using the Stillinger-Weber potential, which 
shows that the [$\overline{2}11$] step on  1$\times$1 reconstructed Si(111) has 
a Schwoebel barrier of 0.61$\pm$0.07 eV, we calculate here the same barrier
corresponding to two types of kinks on this step - one with rebonding between 
upper and lower terrace atoms (type B) and the other without (type A).  
From the binding energy of an adatom, without additional relaxation of other 
atoms, we find that the Schwoebel barrier must be less than 0.39 eV 
(0.62 eV) for the kink of type A (type B). From the true adatom binding 
energy we determine the Schwoebel barrier to be 0.15$\pm$0.07eV 
(0.50$\pm$0.07 eV).  The reduction of the Schwoebel barrier due to the 
presence of rebonding along the step edge or kink site is argued to be a robust
feature.  However, as the true binding energy plots show discontinuities due 
to  significant movement of atoms at the kink site, we speculate on the 
possibility of multi-atom processes having smaller Schwoebel barriers.}} 
\address{\mbox{ }}
\address{\mbox{ }}
\maketitle

             
\narrowtext

\section{Introduction}

The Schwoebel barrier, which was introduced as the additional barrier for 
adatom diffusion over a step edge from the upper to lower terraces, \cite{Shsp}
 has been a subject of current interest for its influence on the growth of a 
singular (flat) surface.\cite{Mdj,MsMp,Awh,Ie,Cjl}  It was pointed out by 
Villain \cite{Vill} that  in the 
presence of such a barrier, growth of by step flows was stable only 
if the surface was sufficiently vicinal with possible instabilities setting in 
during the growth of a flat surface.  It is now accepted that it leads to a 
coarsening of the evolving surface morphology.\cite{Mdj,MsMp,Awh,Ie,Cjl}. 
The present study is motivated by recent observations of another kind of 
instability on the high temperature 1$\times$1 phase of Si(111) - the 
reversible step bunching instability during sublimation. \cite{Avl,EFu}
As these experiments can be reinterpreted in the presence of the Schwoebel 
barrier, \cite{Skod1} we summarize here our previous calculations of the 
same corresponding to straight (high symmetry) steps \cite{Skod1} and 
present results corresponding to kinked steps, both of which use the empirical 
Stillinger-Weber potential. The use of this potential here (and in our previous 
\cite{Skod1}) study has been motivated by the fact that features that follow 
from changes in coordination number (of the adatom probing the potential 
energy topography) are expected to survive even if the details of the 
empirical potential used change. We attempt to identify such features here. 

\bigskip

\begin{figure}
 \vbox to 4.5cm {\vss\hbox to 6cm
 {\hss\
   {\includegraphics{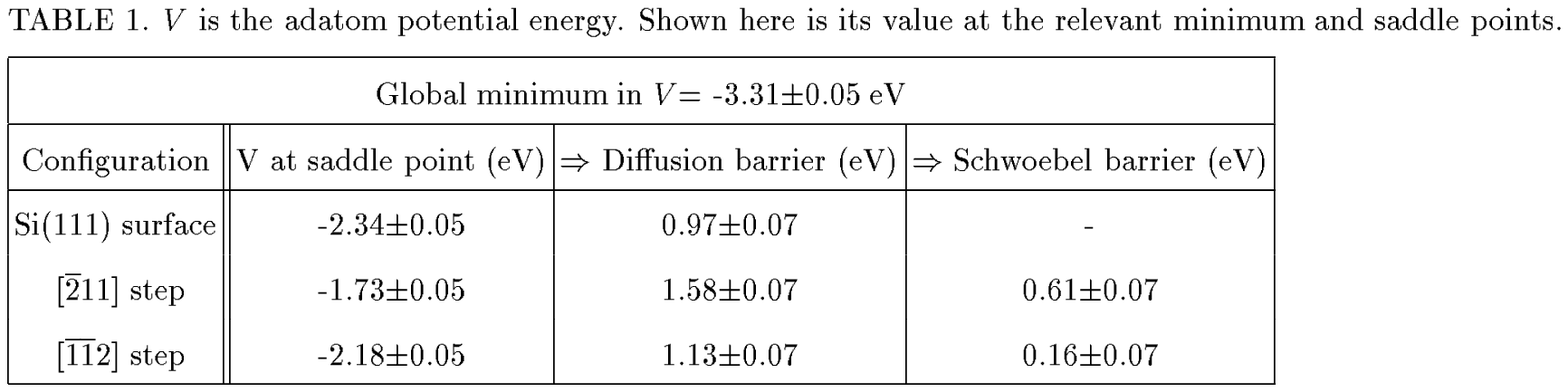}
   }
  \hss}
 }
\end{figure}

Table 1 summarizes our previous results.\cite{Skod1} 
The straight (high symmetry) [$\overline{2}11$] and 
[$\overline{1}$$\overline{1}2$] and other step orientations are shown in Fig.1.
Note that the [$\overline{2}11$] 
step 

\begin{figure}
 \vbox to 8.2cm {\vss\hbox to 6cm
 {\hss\
   {\includegraphics{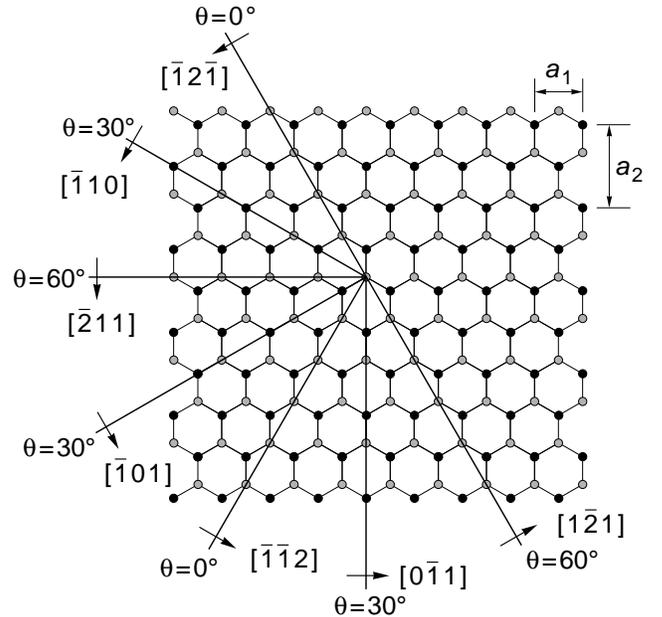}
   }
  \hss}
 }
\caption{One bilayer of the Si(111) surface consisting of the upper monolayer
(grey) and lower monolayer (black). The figure shows the threefold and
reflection symmetry of this surface: Steps running along directions with
equal $\theta$ are identical. }
\end{figure}

\vspace{4.6 cm}

\noindent shows a large Schwoebel barrier of 0.61$\pm$0.07 eV. However, an 
analysis 
of experimental data on the electromigration of steps \cite{EFu} using a 
diffusion equation showed that the upper bound 
on the Schwoebel barrier \cite{Skod1} (in a particular limit of the equation 
parameters) was very small (0.05 eV). Therefore, here we calculate the 
Schwoebel barrier corresponding to unit depth 
kinks on the [$\overline{2}11$] step to determine if it continues to be large.

\begin{figure}
 \vbox to 10.5cm {\vss\hbox to 6cm
 {\hss\
   {\includegraphics{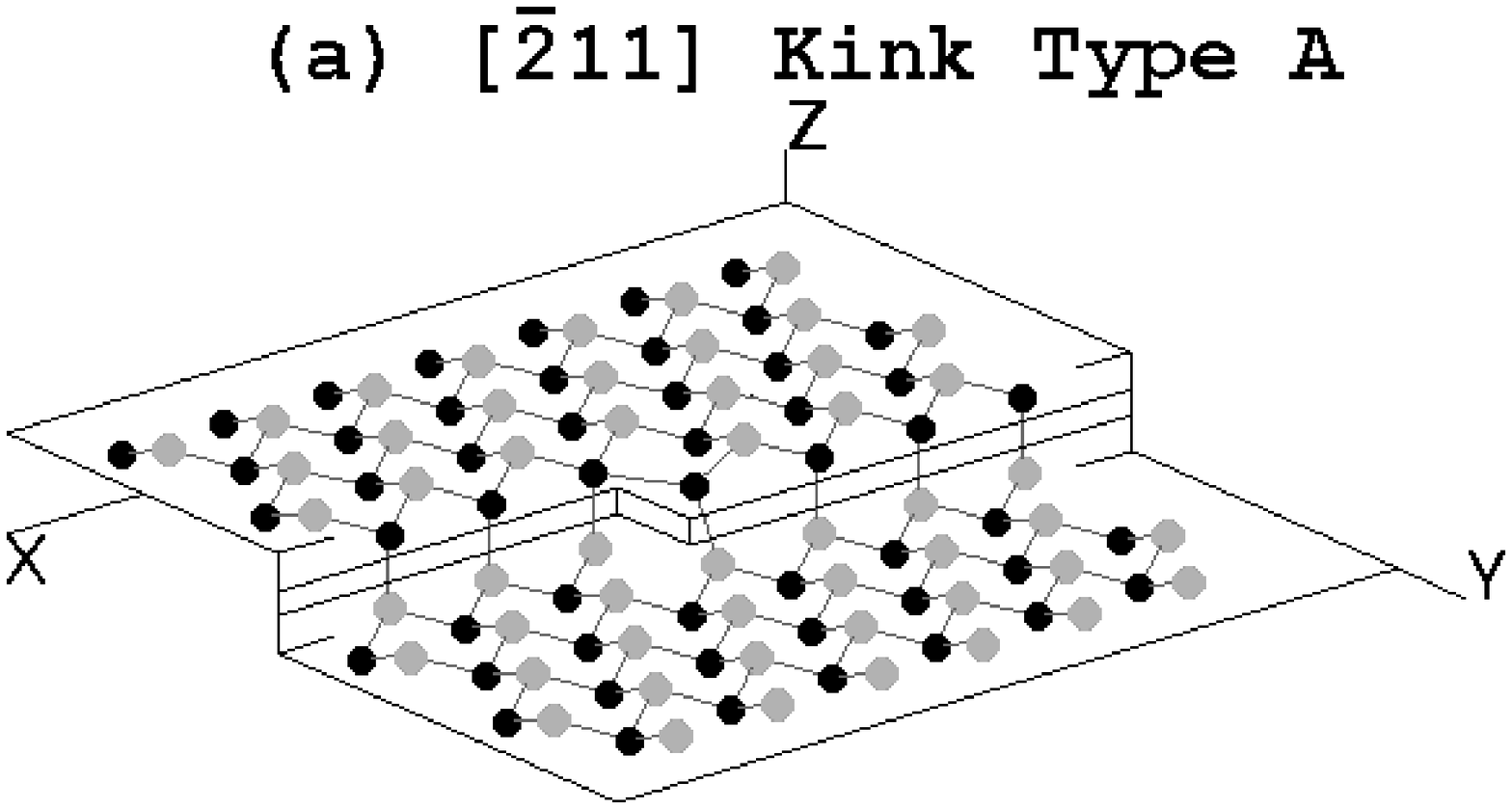}
   }
   {\includegraphics{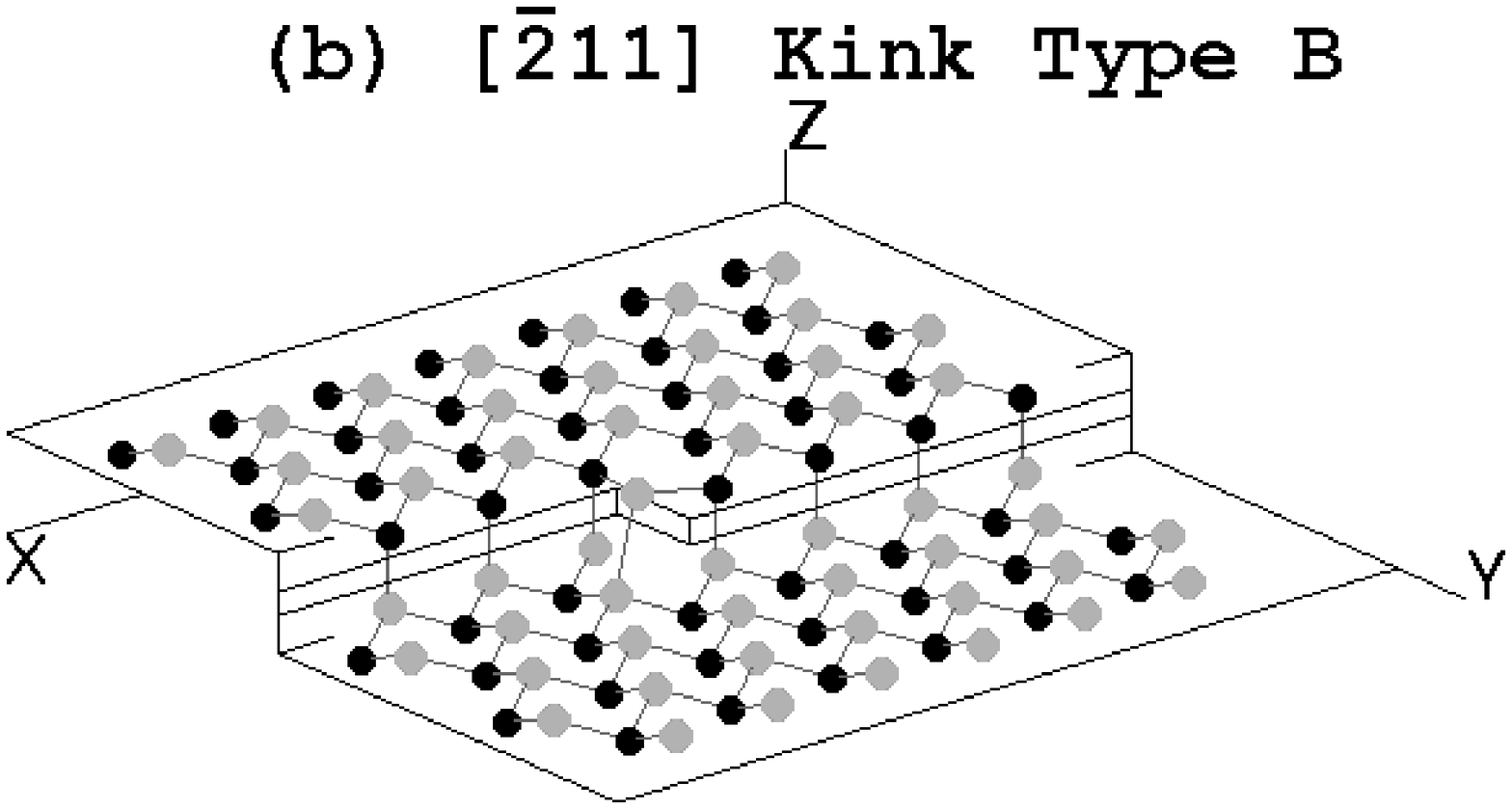}
   }
  \hss}
 }
\caption{The two kinks studied here - type A has atoms rebonded along the kink 
site whereas type B has, at the kink site, an upper terrace atom rebonded to 
the lower terrace.
On both terraces the upper monolayer is shown in grey and the lower monolayer 
in black. }
\end{figure}

\section{Molecular Dynamics Method}

Using the Stillinger-Weber potential, diffusion barriers are determined by 
mapping the adatom potential energy a function of the ($x,y$) position of the 
adatom (in the (111) 
plane) for two types of kinks on the [$\overline{2}11$] step - type A (see 
Fig. 2(a)), which 
has rebonding of atoms along the kink site and type B (see Fig. 2(b)), 
which has, at the kink
site, an upper terrace atom rebonded to the lower terrace. In our previous 
study on straight steps \cite{Skod1}, configurations of type B were neglected 
since the step ([$\overline{1}$$\overline{1}2$]) that allowed for this 
structure would have very large step-step interactions inconsistent with 
experimental estimates.\cite{Skod} However, the large interactions were 
due to the presence of one rebonding atom per lattice constant along the step 
edge. The configuration of type B is nevertheless considered here, since, 
at low kink densities {\it i.e.} for a nearly straight [$\overline{2}11$] step, 
the number of such 
rebonding atoms per unit length along the step edge would be correspondingly 
low and therefore the expected step-step interactions would be 
smaller.
The adatom potential energy $V$ has been computed as the
difference in the minimum potential energy of the system with the adatom at
infinity (non-interacting) and the same with the interacting adatom.

Standard molecular dynamics (MD) procedures of integrating Newton's law (with 
dissipation to reduce temperature) and  the steepest descent equations have 
been used to determine the minimum potential energy of the system. These 
routines determined the adatom potential energy to an accuracy of 
$10^{-4}$ eV. The ($x,y$) 
coordinates of the adatom are fixed during the integration process.
The system consisted (as before \cite{Skod1}) of six bi-layers of Si(111) in 
an MD cell, the bottom 
three layers of which are fixed at bulk lattice coordinates throughout the 
calculation. The system size along the $x$ axis ($\ell_x$) (which was parallel 
the [$\overline{2}11$] step edge) was $5\frac{1}{2}a_1$ 
and along the $y$ axis ($\ell_y$)  was $3\frac{2}{3}a_2$. Periodic boundary 
conditions along the $x$ axis identified the points $(0,0,h)$ ($h$ is the step 
height) and $(0,\ell_y,0)$ and the same along the $y$ axis identified the 
points 
$(0,k,0)$ ($k$ is the kink depth) and $(\ell_x,0,0)$. These boundary 
conditions made it possible to have exactly one vicinal step with one kink 
(and no antikinks) in the MD cell. As before,\cite{Skod1} the
atoms on the ($x,y$) boundaries were however held fixed during the computation 
of $V$ to prevent the
entire configuration from shearing, particularly when the adatom is moved
away from a deep minimum. The kinks were roughly in the middle of the cell 
consisting of movable atoms (see Figs. 3(a) and 4(a)). As the ($x-y$) size 
 of this cell is larger than  that used previously \cite{Skod1} in studying 
the system size dependence of $V$ on the Si(111) surface and since it is 
more square in shape, we expect the error in $V$ due to finite size effects 
to be smaller than before \cite{Skod1} (0.01 eV). It must be additionally
noted that with the system size used here (with the adatom absent), the kink
energy was within 6 meV (4 meV) of that calculated previously \cite{Skod} for 
the kink of type A (type B).
 
The MD procedures began with the initial configurations for each ($x,y$) 
position of the adatom corresponding to the relaxed adatom-free structures. 
The $z$ coordinate of the adatom was then varied in small steps in a wide 
range to roughly determine the point ($z_0$) at which its  potential 
energy is the smallest. Initializing this z coordinate at 
$z_0$, the integration procedures were followed,  first with only the 
adatom relaxing while other atoms remain fixed. The minimum of the 
potential energy reached this way ($V_{lg}$) depends only on the local 
geometry of atoms around the adatom. The atoms that were held fixed are 
then allowed to relax together with the adatom to recompute the minimum which 
is now the true adatom potential energy ($V$). With the kink roughly in the 
middle of the region explored (see Figs. 3(b), 3(c), 4(b), 4(c)), $V$

\begin{figure}
 \vbox to 22cm {\vss\hbox to 6cm
 {\hss\
   {\includegraphics{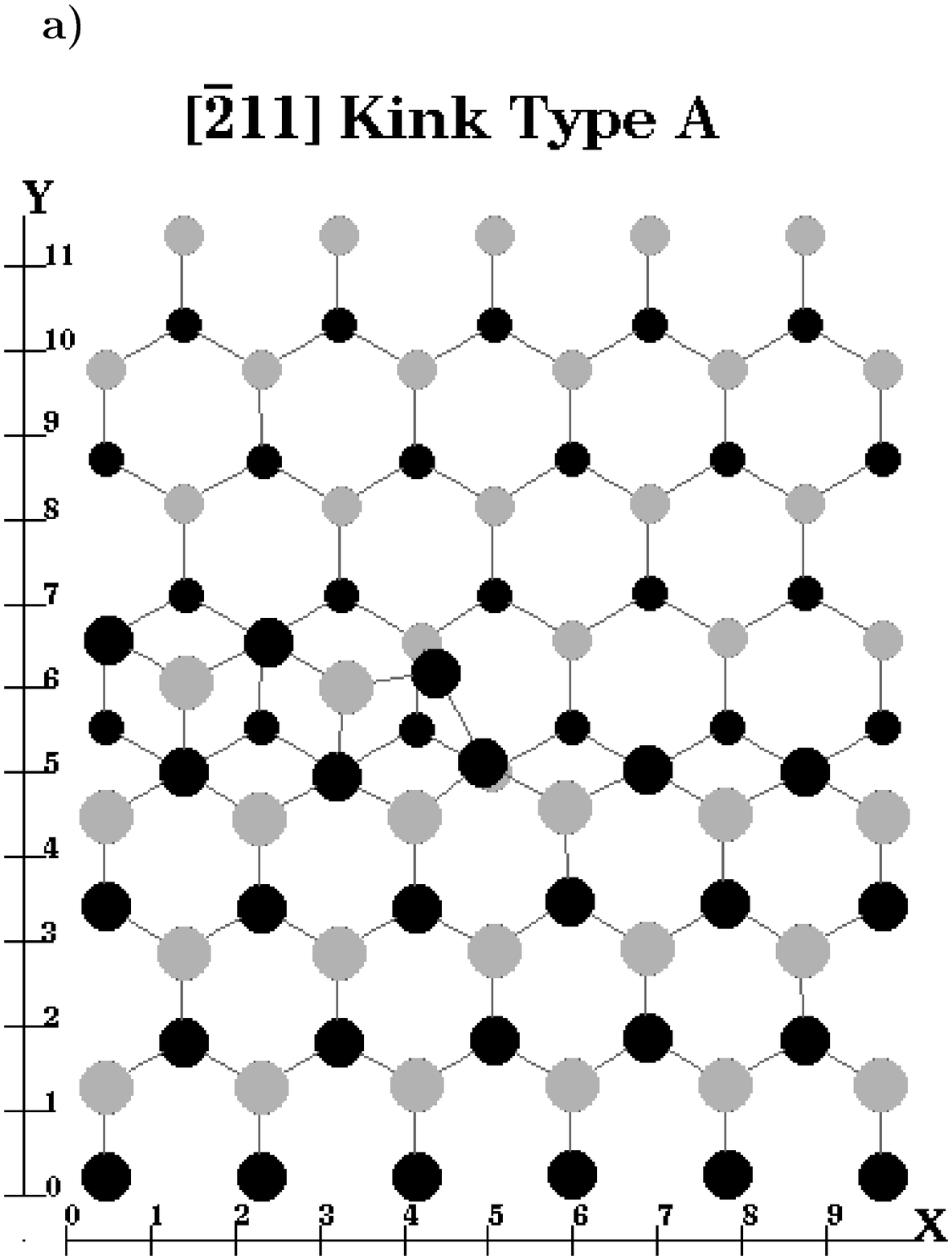}
   }
   {\includegraphics{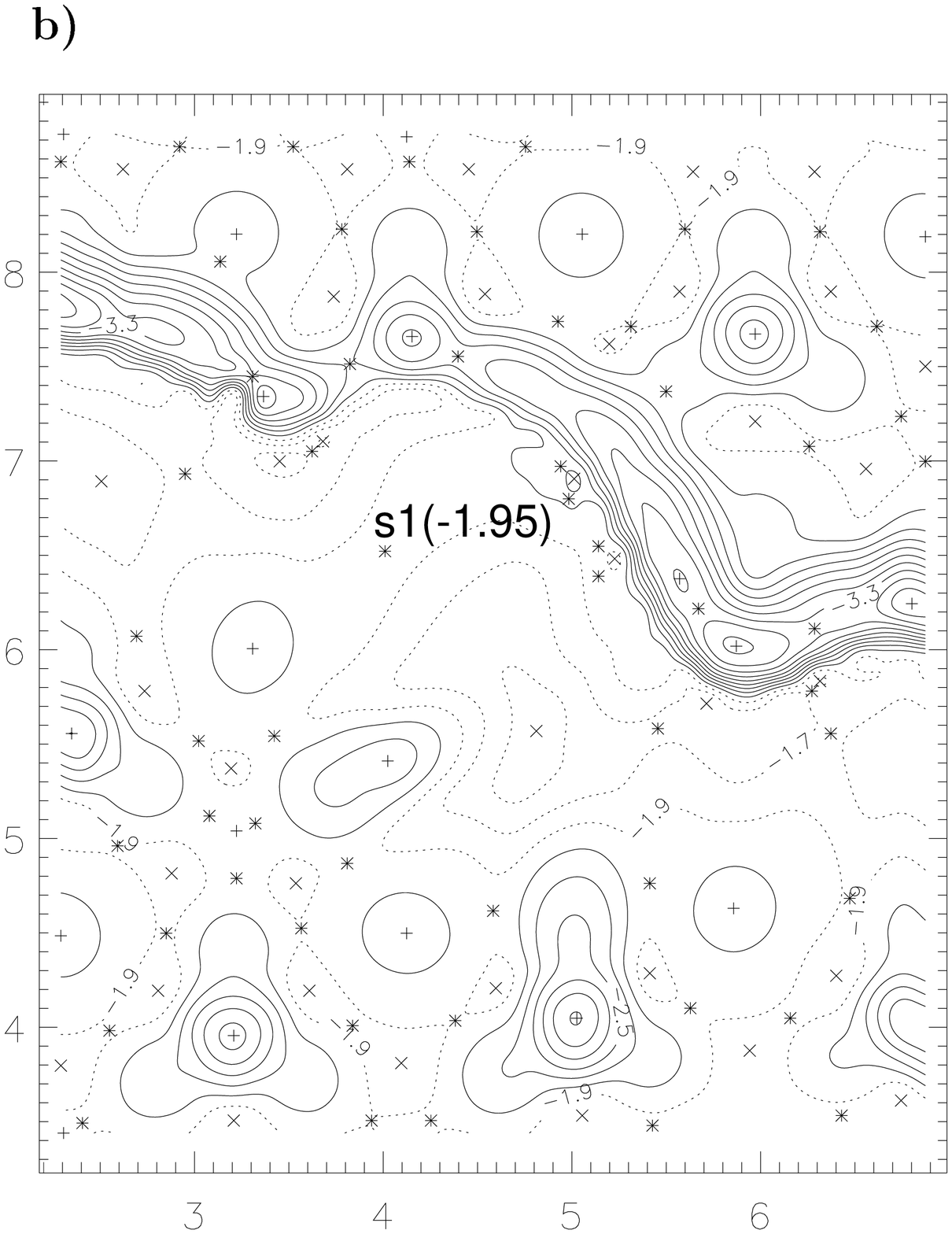}
   }
  \hss}
 }
\end{figure}

\begin{figure}
 \vbox to 11.4cm {\vss\hbox to 6cm
 {\hss\
   {\includegraphics{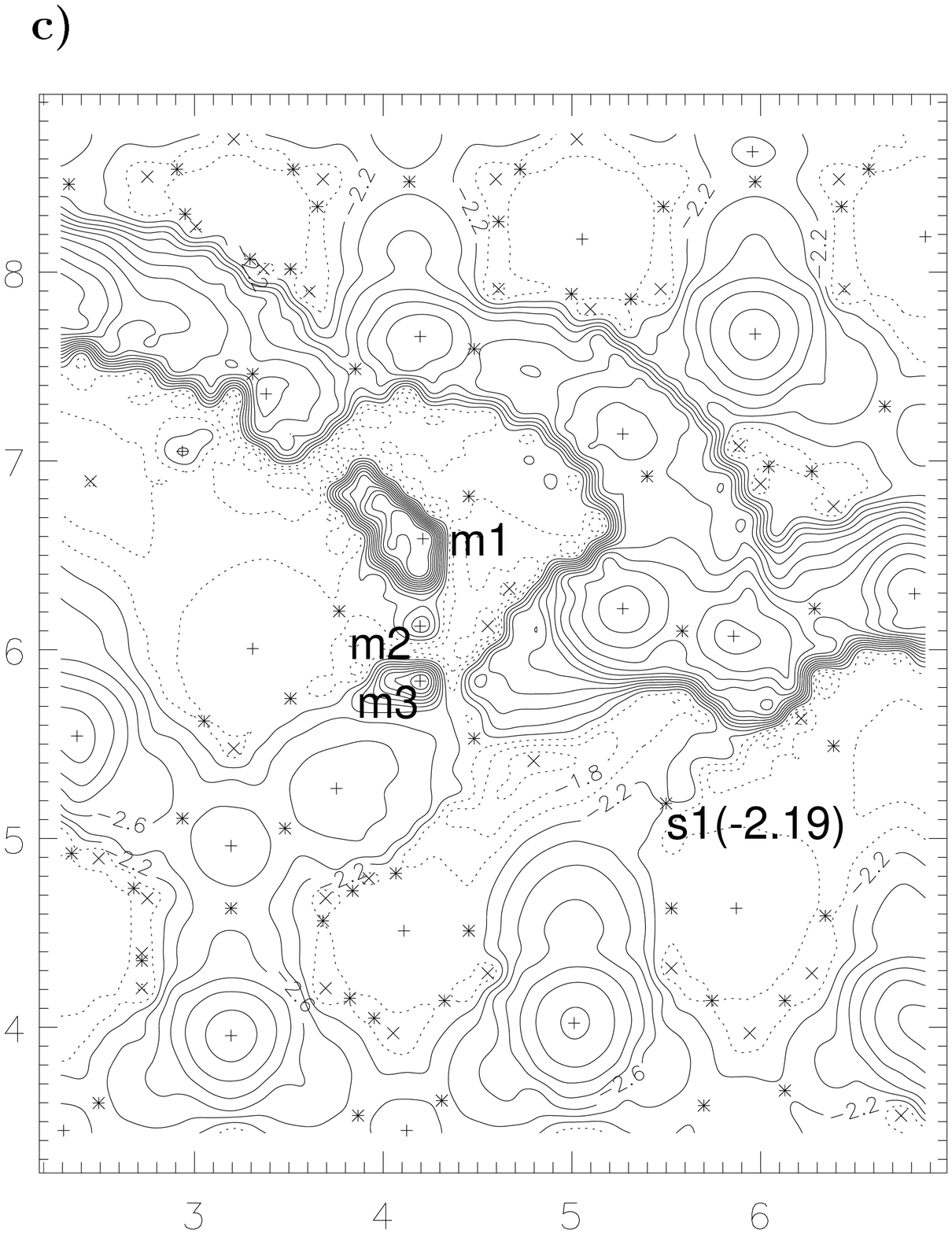}
   }  \hss}
 }
\caption{Shown in (a) is the top view of the kink with rebonding along the kink
site with the upper terrace atoms larger than those on the lower terrace. (b) 
shows the corresponding adatom potential energy derived from the local 
geometry ($V_{lg}$) whereas (c) shows the true adatom potential energy ($V$). 
In both plots, 
contours are separated by 0.2 eV with the minima, saddle points and maxima 
marked (and sometimes labeled) by +(m),*(s) and $\times$(M) respectively. 
Figures in parenthesis are corresponding values in eV.
Contours in (b) $\geq$ -1.9 eV and those in (c) $\geq$ -2.0 eV are marked with 
dashed lines.} 
\end{figure}

\noindent and $V_{lg}$ are computed on a rectangular grid with the spacing 
between 
points being $\frac{a_1}{16}$ ($\frac{a_2}{30}$) along the $x$ ($y$) axis for 
a total length of $2\frac{1}{2}a_1$ ($1\frac{2}{3}a_2$). An interpolation 
scheme applying periodic boundary conditions along the $x$ and $y$ axes similar
 to that used in the simulations, was used to construct the contour plots of 
Fig. 3. From our previous study \cite{Skod1}, the error in $V$ due to 
(the same) 
finite grid size and  similar interpolation scheme was estimated to be 
$\pm$0.05 eV. We therefore assume here that this error remains the same. 
Being much larger than the errors due to finite size effects, it is 
assumed to be the error bar in $V$. Barrier values being differences are 
therefore estimated to have an error bar of $\pm$0.07 eV.

\begin{figure}
 \vbox to 22cm {\vss\hbox to 6cm
 {\hss\
   {\includegraphics{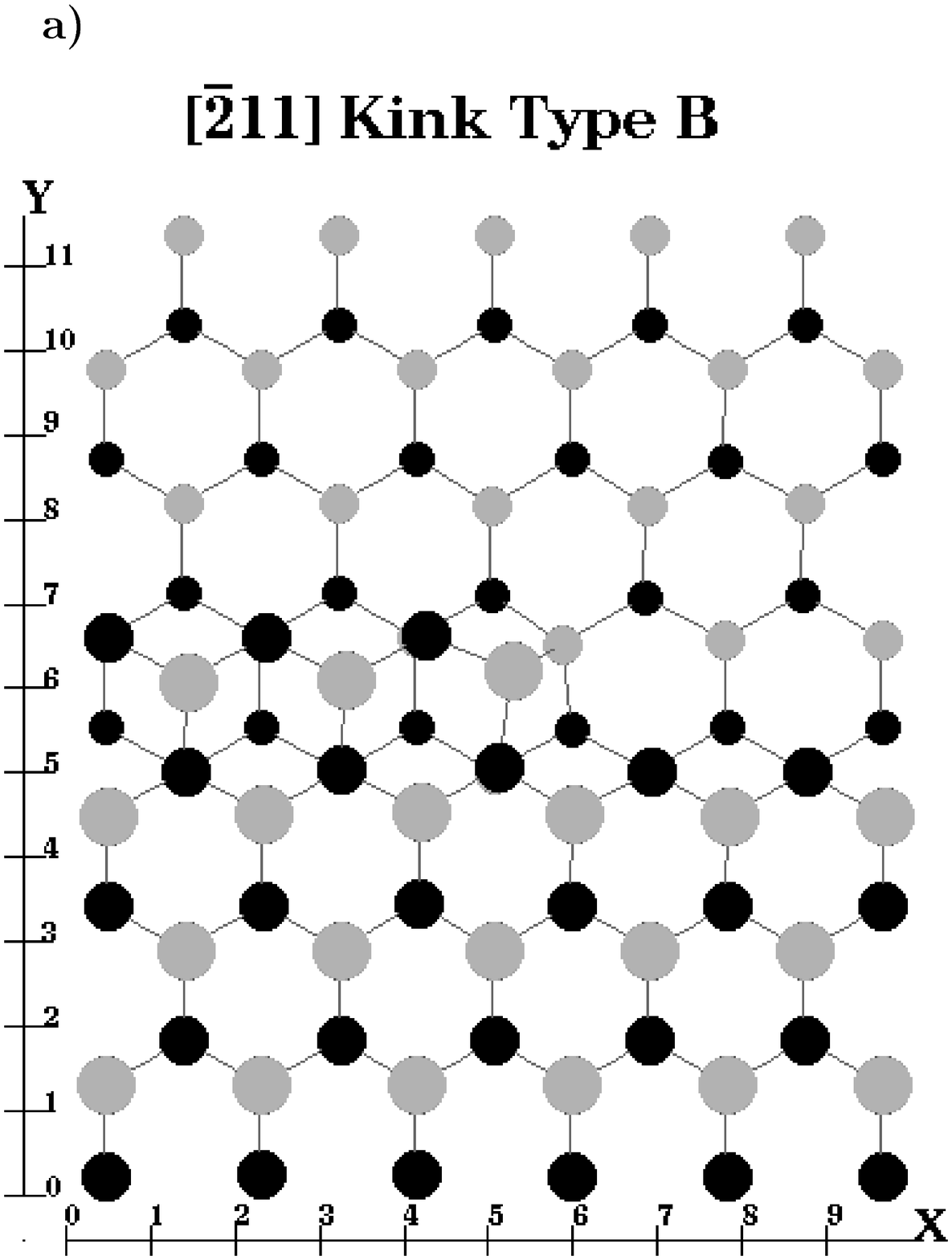}
   }
   {\includegraphics{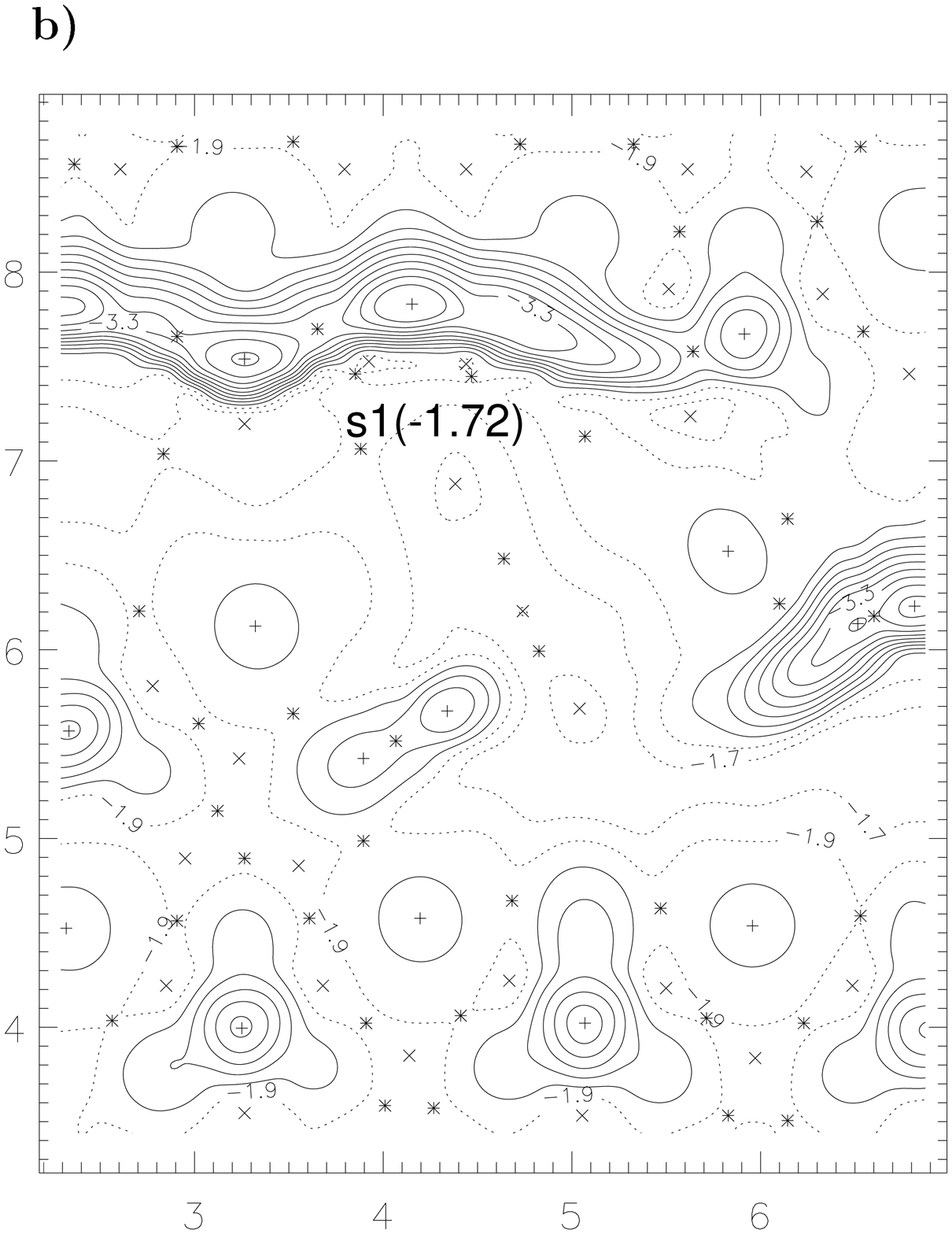}
   }
  \hss}
 }
\end{figure}

\begin{figure}
 \vbox to 11.4cm {\vss\hbox to 6cm
 {\hss\
   {\includegraphics{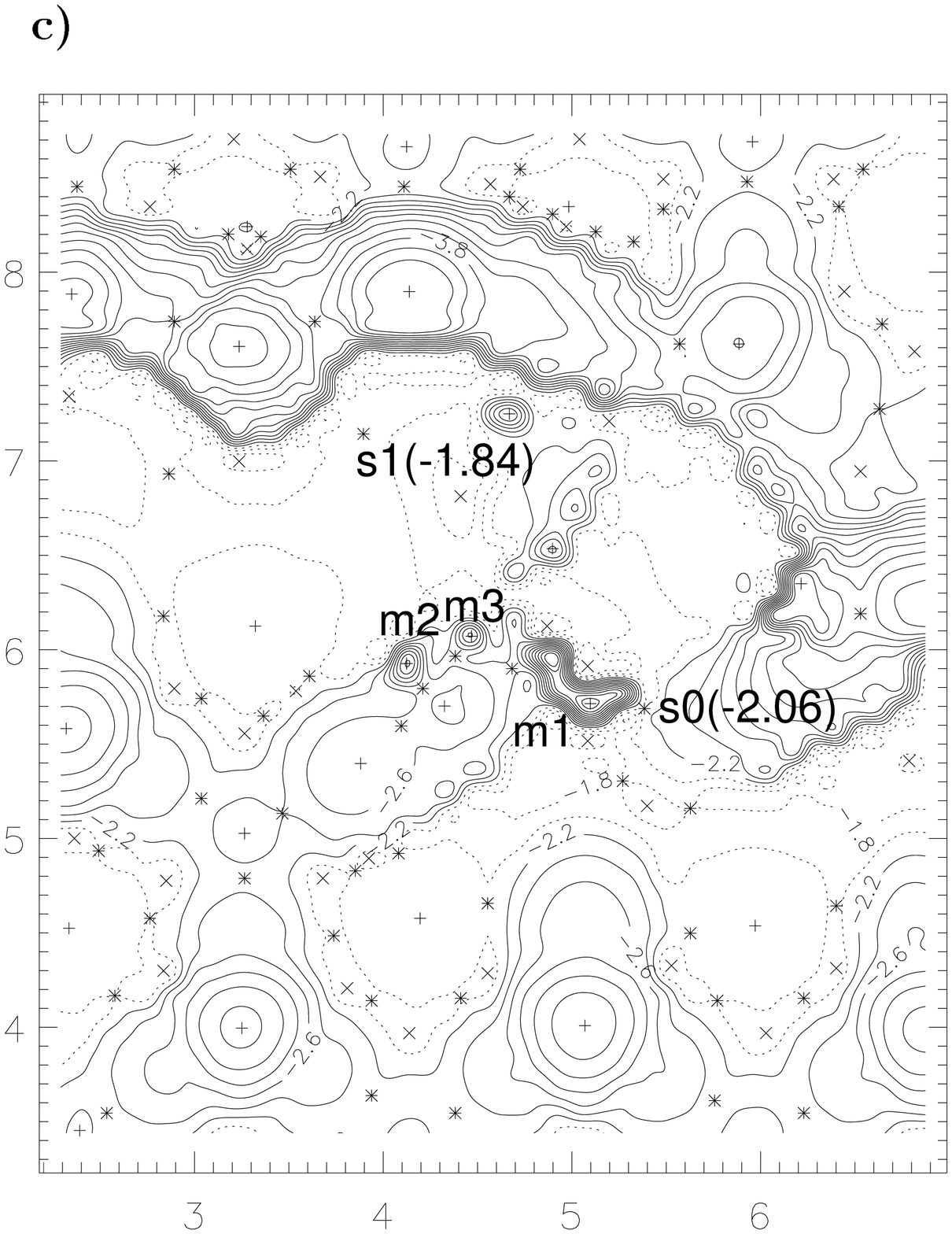}
   }  \hss}
 }
\caption{Shown in (a) is the top view of the kink with an atom of the upper 
terrace (larger atoms) rebonded to the lower terrace. (b)
shows the corresponding adatom potential energy derived from the local
geometry ($V_{lg}$) whereas (c) shows the true adatom potential energy ($V$).
In both plots,
contours are separated by 0.2 eV with the minima, saddle points and maxima
marked (and sometimes labeled) by +(m),*(s) and $\times$(M) respectively.
Figures in parenthesis are corresponding values in eV.
Contours in (b) $\geq$ -1.9 eV and those in (c) $\geq$ -2.0 eV are marked with
dashed lines.}
\end{figure}

\section{Discussion}

As mentioned previously, we have attempted to identify robust features of this 
study as those that follow  from changes in adatom coordination number. Here, 
we assume (as before \cite{Skod1}) that the adatom potential energy obtained 
from relaxing only the adatom over the relaxed (but fixed) adatom free 
configurations ($V_{lg}$) to be a good measure of its coordination number. 
Previously we had argued \cite{Skod1} that features that follow from a 
strong correlation between $V_{lg}$ and $V$ are robust,{\it i.e.}, would 
survive changes in details of the empirical potential used and are expected to 
be reproduced in more satisfactory {\it ab initio} or tight binding 
calculations. Specifically, it was assumed that if the saddle point determining
the Schwoebel barrier in $V_{lg}$  was nearly at the same position as that in 
$V$ then the Schwoebel barrier is a robust feature. We now

\clearpage
\widetext
\begin{figure}
 \vbox to 19cm {\vss\hbox to 6cm
 {\hss\
   {\includegraphics{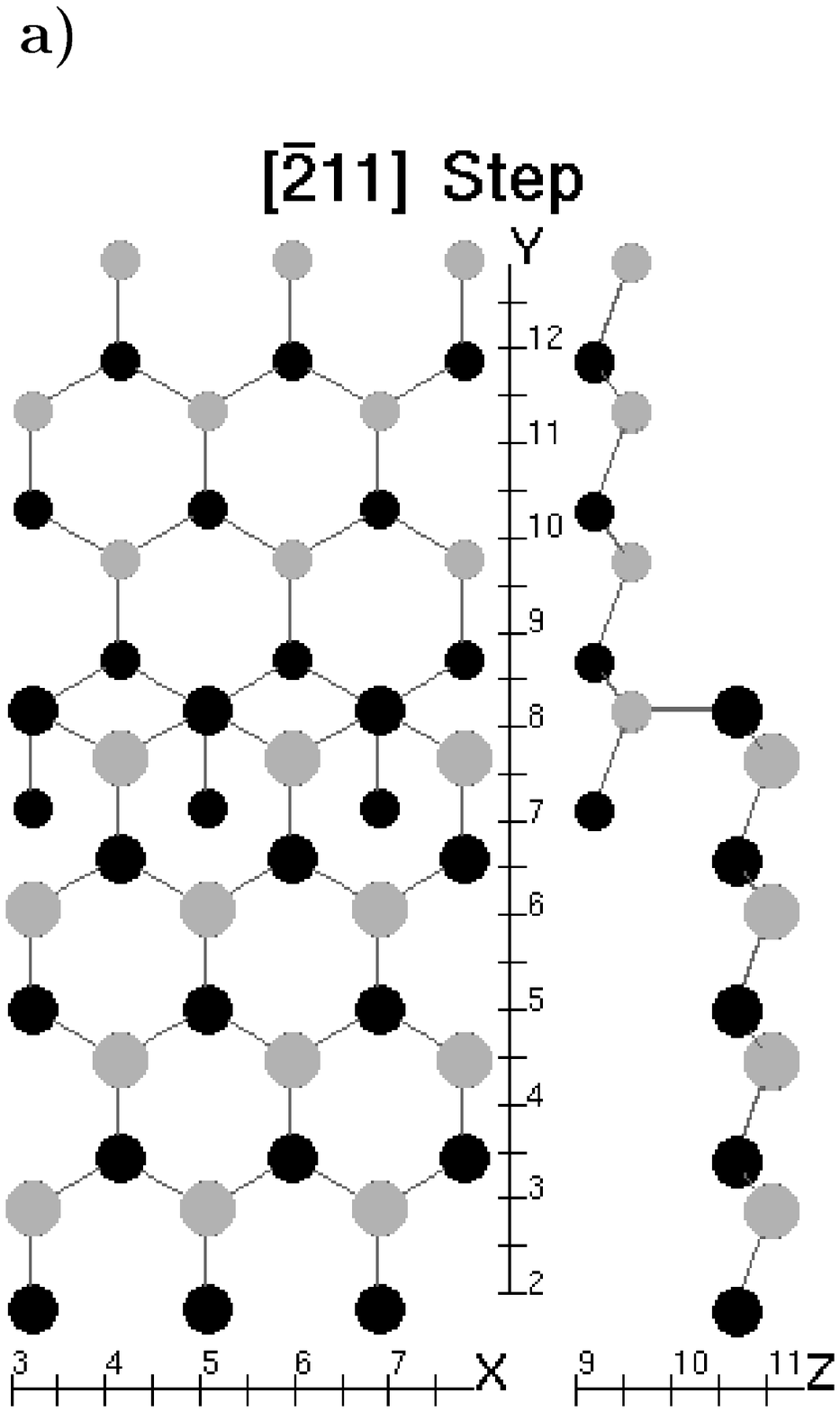}
   }
   {\includegraphics{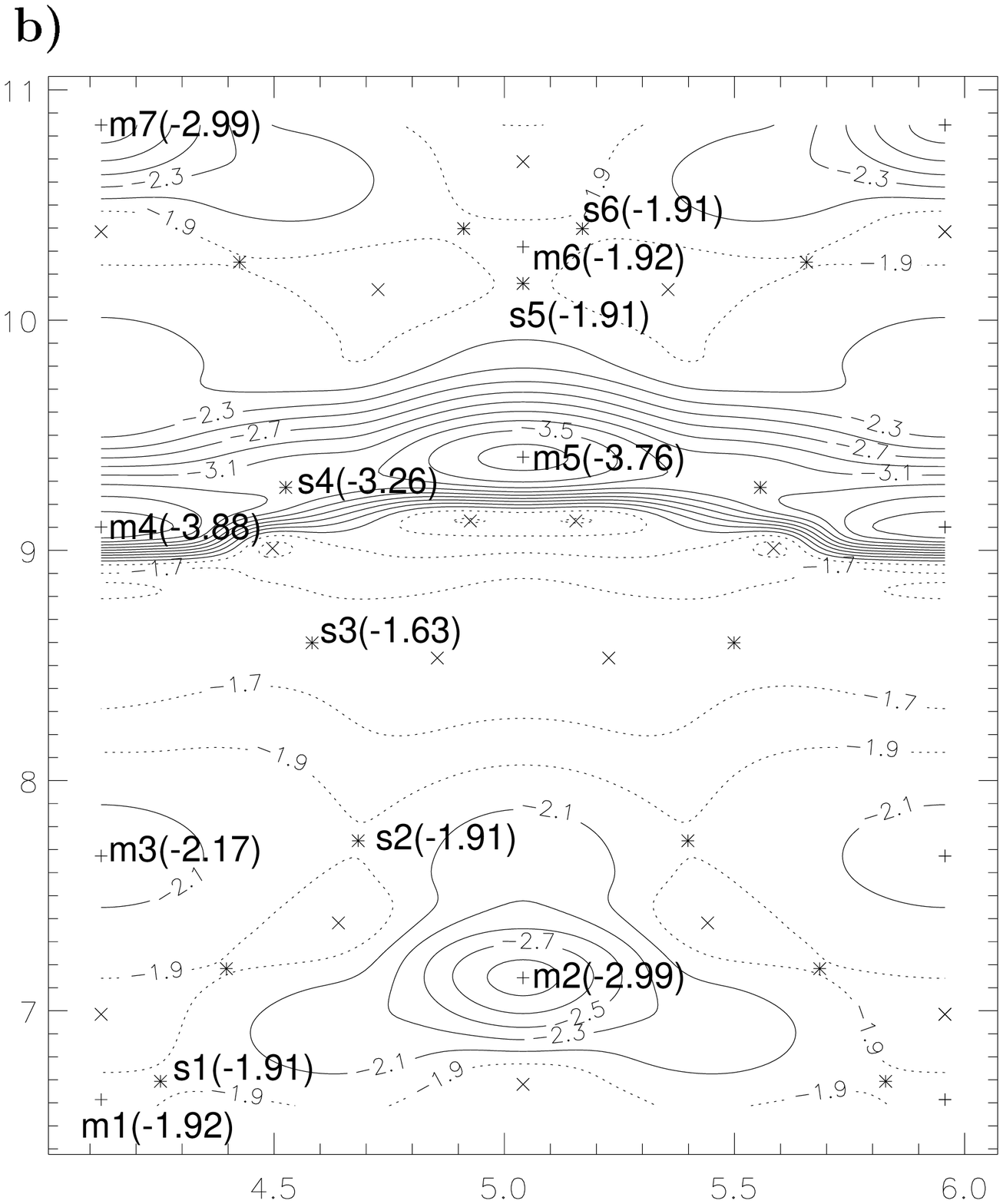}
   }
   {\includegraphics{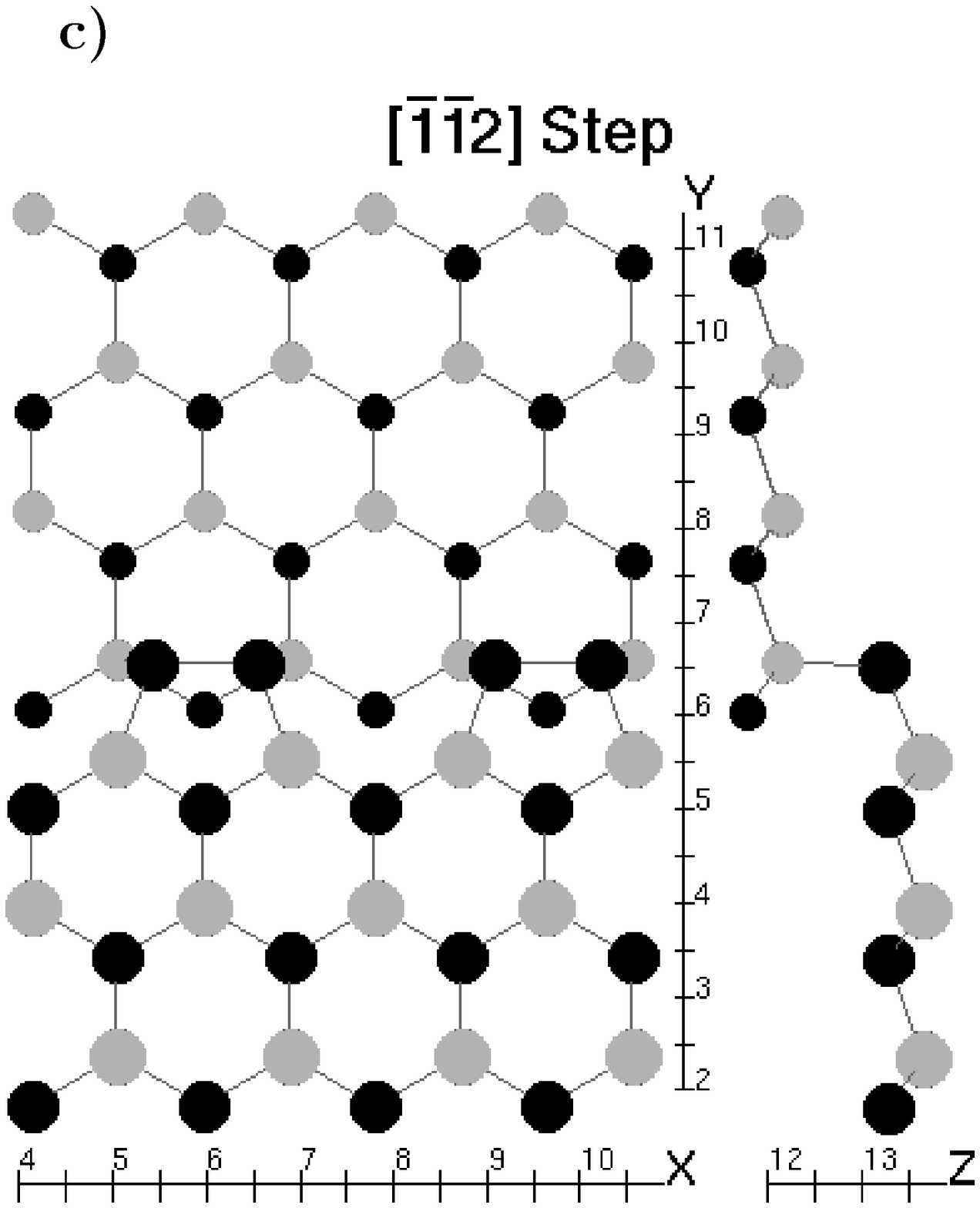}
   }
   {\includegraphics{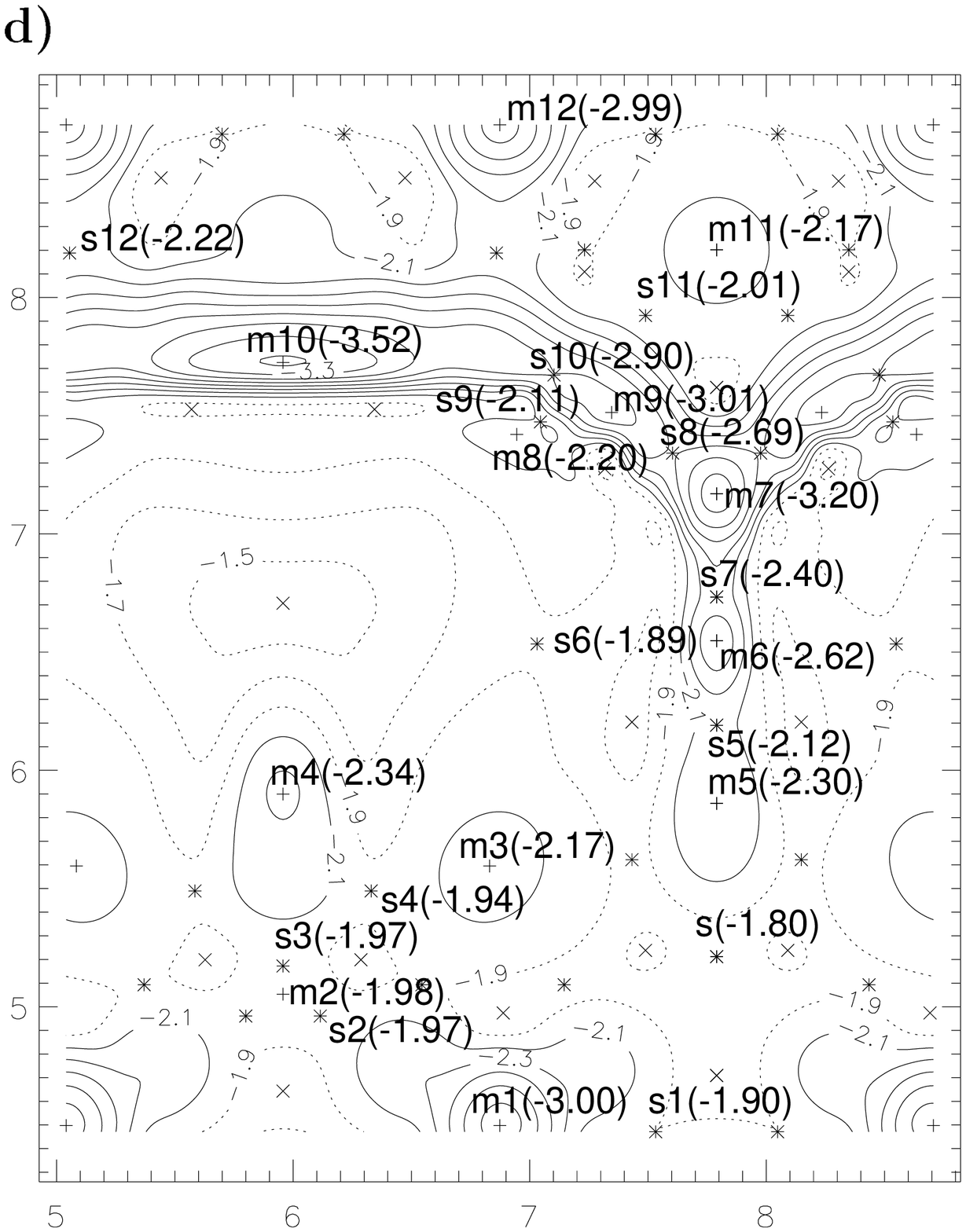}
   }
  \hss}
 }
\caption{ In the step configurations ((a) and (d)) the upper terrace atoms 
are shown larger than those on the lower terrace and within each terrace the 
upper monolayer is in grey and the lower in black. (b) and (d) show the adatom 
potential energy derived from the local geometry ($V_{lg}$) corresponding 
to (a) and (c) respectively. In these plots contours are 0.2 eV apart with 
those $\geq$ -1.9 eV shown by dashed lines. The minima, saddle ponits and 
maxima are  marked (labeled) by +(m), *(s) and $\times$(M) respectively with 
the figure in parenthesis being their corresponding value in eV.} 
\end{figure}

\narrowtext
\noindent  argue that even if 
the position of the relevant saddle points differed, there is however a bound 
on 
the Schwoebel barrier that follows purely from $V_{lg}$ -  this  is the  
difference between $V_{lg}$ at barrier determining saddle point on the step or 
kink configurations and $V$ at the \\

\vspace{20.8 cm}

\noindent diffusion barrier determining saddle point 
on the free Si(111) surface. It is a strict upper bound on the Schwoebel 
barrier (that follows from $V$ on the step or kink configuration) since any 
relaxation that occurs during 
the computation of $V$ can only reduce the relevant saddle point energy. As this
bound is completely independent of the details in $V$, we argue that it must 
be a robust feature.
 
From the contour plots of $V_{lg}$ corresponding to the kinks studied here 
(Figs. 3(b) and 4(b)) as well as the same  corresponding to straight steps 
studied previously \cite{Skod1} (Figs. 5(b) and 5(c)) a common trend emerges - 
the bound 
on the Schwoebel barrier (or equivalently the energy of the barrier determining
 saddle point in $V_{lg}$) is small (or more negative) when atoms at the step 
edge or kink site 
are rebonded. It can be seen (from Figs. 3 and 5) that this saddle point
( $s1$ in Fig. 3(b), $s1$ in Fig. 4(b), 
$s3$ in Fig. 5(b) and $s6$ in Fig. 5(d)) is in the neighborhood of an 
upper terrace atom that has moved from its bulk terminated position in a 
direction 
away from the saddle point due to the presence of rebonding at the step edge 
or kink site. The magnitude of this displacement (in the $x-y$ plane) is found 
to be correlated to 
the saddle point energy which is more negative if the displacement is large.  
The displacement (and the relevant saddle point energy) for the 
[$\overline{2}11$] kinks of types A, B, the [$\overline{2}11$] step and the 
[$\overline{1}\overline{1}2$] step is 0.99$\AA$ (-1.95 eV), 0.21$\AA$ 
(-1.72 eV), 0.00$\AA$ (-1.63 eV) and 0.72$\AA$ (-1.89 eV) respectively. 
Therefore the bound on the corresponding Schwoebel barrier is 0.39 eV, 0.62 eV,
0.71 eV and 0.45 eV respectively. 

The true adatom potential energy $V$, corresponding to the kink configurations 
studied here, is shown in Figs. 3(c) and 4(c). These plots show 
discontinuities near the kink site indicated by the the presence of minima, 
some of which are labeled in the figures. Studying the final atomic 
configuration when the adatom is in these regions shows that the 
discontinuities are due to large 
scale rearrangements of atoms leading to a loss in the identity of the adatom.
In other words, in these configurations, the adatom seems to occupy a lattice 
position after dislodging another atom which now appears to be the adatom. 
Diffusion in these regions cannot therefore be viewed as a $single$ atom 
process, therby making the  contour plots in these regions less meaningful. 
The saddle point determining the Schwoebel barrier in case of the 
[$\overline{2}11$] kink of type A ($s1$ in Fig. 3(c)) is however not in the 
proximity of such regions. Therefore the corresponding Schwoebel barrier of
0.15$\pm$0.07 eV is assumed to be relevant for processes involving only 
a single addtom. In the case of the kink of type B, it appears that $s0$ (in 
Fig. 4(c)) is the relevant saddle point that determines the Schwoebel barrier. 
However, this point is close to a minimum 
that indicates a discontinuity in $V$. The Schwoebel barrier determined by 
this point (0.28$\pm$0.07 eV) is therefore considered as corresponding to a 
multi atom, and not a 
single atom, process. Hence, discounting this point, the Schwoebel barrier 
(determined by $s1$ in Fig. 4(c)) is 0.50$\pm$0.07 eV. 

In our previous study of barriers over straight steps, \cite{Skod1} 
discontinuities in $V$, of the kind that are seen here, were not observed.
This was because the temperature used during the simulation was very small - 
$\approx 2\times10^{-4}$ eV. Here we used a larger temperature 
($\approx 3\times10^{-3}$ eV) which resulted, in the presence of the adatom, 
the large scale movement of atoms around the kink site. The experimentally
relevant temperature \cite{EFu} however continues to be much larger - 
$\approx$ 0.1 eV. 
At these temperatures we expect multi-atom processes to have smaller 
Schwoebel barriers - this is supported by the observation here that such 
a process occurs near the [$\overline{2}11$] kink of type B with the Schwoebel 
barrier being smaller than that for the single atom process. The general 
decrease in the barrier values is however consistent with the small upper 
bound on the
Schwoebel barrier (0.05 eV) developed previously \cite{Skod1} from an analysis 
of experimental data on the electromigration of steps. \cite{EFu} 

\section{Conclusion}
Schwoebel barriers, calculated using the empirical Stillinger-Weber potential, 
for unit depth kinks along the [$\overline{2}11$] step are smaller in 
magnitude as compared to that calculated previously \cite{Skod1} for the 
straight [$\overline{2}11$] step. This decrease can be expected directly form 
the adatom potential energy plots ($V_{lg}$) that follow purely from the local 
geometry 
of atoms around the adatom in these atomic configurations. These plots as 
well as similar plots for the straight [$\overline{2}11$] and 
[$\overline{1}\overline{1}2$] steps calculated previously \cite{Skod1}
show that the upper bound on the true Schwoebel barrier calculated using 
$V_{lg}$ is correlated to the 
the displacement of an atom on the upper terrace of these configurations that 
is near the relevant saddle point in $V_{lg}$. The true adatom potential energy
plots ($V$) however show discontinuities due to large scale movement of 
atoms near the kink sites resulting sometimes in smaller barriers than 
when such movements do not occur. We therefore speculate that multi-atom 
processes occurring in  the configurations studied here and previously 
\cite{Skod1} could have smaller Schwoebel barriers.

\section{Acknowledgments}   
This work has been supported by the NSF-MRG and the U.S. ONR.

\end{document}